\documentclass[12pt]{article}      
\usepackage{amsmath,amssymb,amsfonts} 
\usepackage{graphics}                 
\usepackage{color}                    
\usepackage{hyperref}                 
\usepackage{latexsym}
\usepackage{verbatim}
\usepackage{graphicx}

\begin{document}

\vspace{20pt}

\begin{center}

{\LARGE \bf Some examples of F and D-term SUSY breaking models} \\[1.5ex]

\vspace{30pt}

{\bf  Luis Matos}

{\small \em {Institute for Particle Physics Phenomenology, \\
Department of Physics, Durham
University,\\ Durham DH1 3LE, United Kingdom}}

\vspace{10pt}

{\sffamily \tt
l.f.p.t.matos@durham.ac.uk}

\vspace{30pt}
\end{center}

\begin{abstract}
In this paper we investigate the possibility of finding models that, in the global minimum, break SUSY by a combined F and D-term effect. We find that if the superpotential is a cubic polynomial in the fields and the Kahler potential is canonical this does not happen.

We then give two examples where the previous conditions do not apply and where the vacuum of the model breaks SUSY by combined F and D-term vevs:
SQCD+singlet with $N_C=N_F+1$, and a baryon deformed ISS model.

We briefly discuss the phenomenological viability of the models.
\end{abstract}

\setcounter{page}{0}
\thispagestyle{empty}
\newpage

\section{Introduction}

\paragraph{}

There has been, after the construction of Intriligator, Seiberg and Shih \cite{Intriligator:2006dd}, an interest in studying models that break SUSY on a metastable ground state. Many of these consist in deformations of the original ISS model in such a way that R-symmetry is either approximate\cite{Murayama:2007fe,Murayama:2006yf} or spontaneously broken\cite{Abel:2007jx,Abel:2007nr,Abel:2008gv}. The breaking of R-symmetry is important if upon integration of the messenger fields we want to generate Majorana masses for the supersymmetric standard model (SSM) gauginos.

In the IR, most of these models can be described by some O'Rafeartaigh-like dynamics. Supersymmetry is broken because not all the F-term equations can be solved, often appealing to the rank condition.  In many of these, unfortunately, gauginos were predicted to be much lighter than scalars.  This is not so much related to how one breaks R-symmetry, but because we are expanding the potential around its tree level minimum \cite{Komargodski:2009jf}.

The reason why it's not good to have a large hierarchy between these masses has to do with the amount of fine-tuning one faces when trying to include a mechanism that generates $\mu$ and $B_{\mu}$ terms to achieve electroweak symmetry breaking.

The understanding of why the ratio of gaugino and scalar masses is large in these models already gives some hints of ways to avoid it. One possibility \cite{Giveon:2009yu}  is to use the ISS construction, but expand the theory around a different vacuum which has a higher energy. One has to ensure that the vacuum is stabilized somehow, but the argument of the vanishing of gaugino masses does not apply. A similar possibility is of a hybrid model that mixes explicit and direct mediation in such a way that one can tune the parameters of the theory to decide how metastable the vacuum is \cite{Abel:2009ze} and change the gaugino/scalar mass ratio. 
This approach assumes a different kind of metastability than the one of the original ISS proposal, as in these models there is a SUSY vacuum that can be found within the limits of validity of the effective low energy theory.

Another hypothesis, which we will follow here, is to consider the effects of switching on the gauge coupling and look for a combined effect of F and D-term SUSY breaking. Examples were found in the context of semi-direct mediation in \cite{Seiberg:2008qj,Elvang:2009gk}, or \cite{Raby:1998bg,Dienes:2008rg}. 

These theories of semi-direct mediation also have problems in generating large gaugino masses when compared with scalar masses. An approach to solve this problem has been given in\cite{Ibe:2009bh}.

One could also try to look for pure D-term SUSY breaking by introducing a Fayet-Iliopolous term. However \cite{Komargodski:2009pc} this is not preferred. This is because such a term is not invariant under complexified gauge transformations. Under these, the F.I. term induces a Kahler transformation. This is a kind of transformation that has no impact on the lagrangian of the system, but makes the supercurrent multiplet not gauge invariant. So, if such a term is not present in the high energy theory, gauge invariance of the supercurrent protects it from being generated at low energies. If it is there at high energies, this theory can only be coupled to supergravity if there is an exact global R-symmetry(in the SUGRA theory). It is, however, problematic to have a theory of gravity with a global symmetry, so F.I. terms should vanish. This argument does not apply to ``field dependent F.I. terms'' coming from ``anomalous`` U(1)'s \cite{1987NuPhB.289..589D}, but only true (field independent) F.I. terms.

The other question that arises is the why should the universe be in a metastable vacuum. One possible reason for this is that because of finite temperature corrections to the effective potential in the early universe would make this state lower in energy and thus the true vacuum of the theory\cite{Abel:2006my,Katz:2009gh}.

In this paper we will look at the possibility of having models of combined F and D term SUSY breaking on the global minimum of the (effective) theory. In section 2 we will give some constraints on classes of models that can break SUSY in this way. In Section 3 we will study the case of SQCD with $N_C>N_F$ and a singlet, to see that the vacuum breaks SUSY with both F and D-terms. This example is, in some sense, complementary to \cite{Izawa:1996pk,Intriligator:1996pu} that study SUSY breaking in models with vector-like matter and evade the Witten index argument. However, the minimum we find does not have any flat direction. In section 4 we will study a baryon deformed ISS model and see that the stabilization of the runaway by the Coleman-Weinberg contribution generates small D-terms; we also comment on the phenomenological viability of this model. In section 5 we present our conclusions.

\section{Some constraints on models that can have combined F and D-term SUSY breaking}

In this section we will show that if the superpotential is renormalizable and the Kahler potential is canonical, the model can only break SUSY through F-terms (globally), i.e. when minimizing the function $V_D+V_F$ one always finds $V_D=0$. To do this we will find a set of solutions that minimizes $V_F$ and then show that it's always possible to chose one that solves $V_D=0$. Since $V_D \geq 0$ this is a global minimum of the model.

Since we want to minimize $V_F$, we start by reviewing some known results about O'Raifeartaigh (O'R.) models\cite{Ray:2006wk,Komargodski:2009jf}: 

Let us consider a renormalizable Wess-Zumino model with superpotential W (i.e. W is a degree 3 polynomial in the fields), the Kahler potential is canonical. Then, $V_F=\sum_i|W_i|^2$, where $W_i$ is the derivative of the superpotential with respect to the field $\phi^i$ and indices are raised and lowered by complex conjugation. 

Since we are assuming that the minimum of the potential breaks SUSY, not all the $W_i$ can be made to vanish simultaneously. At the minimum, the gradient of the potential must vanish, $W_{ij}W^j=0$ (i.e. we are not considering cases where there are runaway directions, even though they might be interesting).

The tree-level boson mass is:
\begin{equation}
\label{bosonmass}
m_{B}^2=\left(
\begin{array}{c c}  
\mathcal{M}_F^* \mathcal{M}_F & \mathcal{F}^*\\
 \mathcal{F} & \mathcal{M}_F \mathcal{M}_F^*\\
 \end{array}\right)
\end{equation}
where $\mathcal{F}_{ij}=W_{ijk}W^k$, and $\mathcal{M}_F=W_{ij}$. In a consistent vacuum $M_B^2$ must be positive semi-definite.

We will now see that in the direction $(W^i, W_i)$ this scalar mass matrix has a zero eigenvalue:

\begin{equation}
  \left(
\begin{array}{c}
W_i \\
W^j\\
\end{array}
\right)^{\dagger}
\left(
\begin{array}{c c}  
\mathcal{M}_F^* \mathcal{M}_F & \mathcal{F}^*\\
 \mathcal{F} & \mathcal{M}_F \mathcal{M}_F^*
 \end{array}
\right)
\left(
\begin{array}{c}
W^i \\
W_j
\end{array}
\right)
=2 Re(W^i \mathcal{F}_{ij}W^j).
\end{equation}
Since this mass matrix is positive semi-definite, this must vanish, otherwise it could be made negative by rotating the phase of $W^i$. 
So $W^i \mathcal{F}_{ij}W^j=0$. It's possible to show, by performing an expansion of the potential to 3rd order, and using this equation, that one actually has the stronger result: $\mathcal{F}_{ij}W^j=0$. 

If the superpotential is renormalizable, then the model has a classically flat direction parametrized by $\phi^i=(\phi^i)^{(0)}+z W^i$:
\begin{equation}
\label{NoFtermchange}
 \delta W_i= W_{ij} zW^j+W_{ijk}(z W^j)(z W^k)=0
\end{equation}
This expansion is exact since higher derivatives of W vanish. So, the F-terms are constant along this direction.

We will now consider the two possible situations:
\begin{enumerate}
 \item The F-terms do not break gauge symmetry;
\item The F-terms break gauge symmetry;
\end{enumerate}

The first option is, for example, the case of the ISS model, the second case looks like the 3-2 or 4-1 models, except that in those cases the superpotentials are not 3rd order polynomials.

In the first case the non-vanishing F-terms are all gauge invariant. This means that if we perform a complex gauge transformation on the vevs of the fields, the potential remains invariant. In this case the symmetry group under which the potential $V_F$ is invariant is enlarged from SU(N) to SL(N). This symmetry is enough to solve all the D-term equations. 

One should note that this is not a true complex gauge transformation as only the vevs of the fields, and not the fields themselves, are being rescaled. Meaning, if we were performing a complexified gauge transformation, invariance of the potential would be guaranteed by a shift in the lowest component of the vector superfield. In the case we are considering, the SL(N) symmetry of the vacuum would still be there if there was no vector superfield (i.e. gauge symmetry). 

We consider two cases to illustrate the point: one explicit example and a general argument.

Take a ISS model where we have gauged a U(1) baryon symmetry. The model has gauge group U(1) and the flavour group is SU(6) (for this number of flavours the magnetic gauge group of the ISS is empty). The field content is: quarks, $\phi$, that transform under the fundamental of color and flavour, antiquarks, $\tilde{\phi}$, that transform as anti-fundamentals of color and flavour, and some mesons, $\Phi$, that are color singlets and transform under the adjoint plus a singlet of flavour. 

The superpotential is:
\begin{equation}
 W= h(Tr[\tilde{\phi} \phi \Phi ]-\mu^2 Tr[\Phi])
\end{equation}

SUSY is broken by the rank condition: 
\begin{equation}
 F_{\Phi_{ij}}=h(\tilde{\phi}_j \phi_i-\mu^2 \delta_{ij})
\end{equation} 
Since the number of colours is less than the number of flavours, it's not possible to solve all these equations and SUSY is broken. The scalar potential is then minimized to be:
\begin{equation}
 V_F= 5h^2\mu^4
\end{equation}

The moduli space up to global symmetries is given by:
\begin{equation}
 \Phi=\left(\begin{array}{cc}
  0 & 0\\
  0 & X\\
            \end{array}\right), \phi=\left(\begin{array}{c}
 \phi_0 \\
0\\
\end{array}\right),\tilde{\phi}=\left(\begin{array}{c}
 \tilde{\phi}_0 \\
0\\
\end{array}\right),
\text{with} \quad\tilde{\phi}_0 \phi_0=\mu^2
\end{equation}

The D-term potential in a random point of the pseudomoduli space is
\begin{equation}
 V_D= \frac{g^2}{2}(|\phi_0|^2-|\tilde{\phi}_0|^2)^2
\end{equation}

Under a complexified gauge transformation of the vevs, 
\begin{equation}
\begin{array}{c}
 \phi \rightarrow e^{ \alpha}\phi\\
\tilde{\phi}\rightarrow \tilde{\phi}e^{-\alpha}\\
\end{array}
\end{equation}

And $V_D(\alpha)=  (e^{2\alpha}|\phi_0|^2-e^{-2\alpha}|\tilde{\phi}_0|^2)^2$. This vanishes when $\alpha=\frac{1}{4} Log[\frac{|\tilde{\phi}_0|^2|}{|\phi_0|^2}$. Note that on the F-term moduli space, neither $\phi$ or $\tilde{\phi}$ can vanish.

For the second example we just recall the argument given in\cite{Wess:1992} and note that it still applies if the F-terms are non-vanishing, but gauge invariant.

For this particular example $SU(N)$ is semi-simple, so it has a Cartan sub-algebra. The D-terms transform in the adjoint representation of the group, and it is always possible to find a group transformation that rotates any initial D-term into the direction given by one particular element ( call it T) of the Cartan subalgebra. This gauge transformation gives another (gauge equivalent) solution to the minimization of the potential $V_F$. 

This generator $T=diag(\{\mu_i\})$ (where $diag(\{\mu_i\})$ is a diagonal matrix with eigenvalues $\{\mu_i\}$), and
\begin{equation}
 \tilde{D}=\phi^{\dagger}.diag(\{\mu_i\}).\phi
\end{equation}
Where $\phi$ and dot product is used as a notation for the sum over all the fields with the generator in the appropriate representation.

Because the F-terms are gauge invariant, we can actually complexify the parameter that defines the gauge transformations, and still have a (physically distinct) solution of the minimization of $V_F$. This argument only differs from \cite{Wess:1992} on the assumption that all the F-terms vanish.

We now distinguish two cases:
\begin{itemize}
\item all $\mu_i$ have the same sign.
\item the $\mu_i$ take both signs
\end{itemize}

Under a complexified gauge transformation the D-term changes as
\begin{equation}
 \tilde{D}=\phi_i \mu_i e^{2 \mu_i \eta}\phi^i
\end{equation}

In the first case we just need to take $\eta \rightarrow -\infty$ if all the  $\mu_i>0$ or $\eta \rightarrow +\infty$ if $\mu_i<0$, to have a solution to $V_D=0$.

In the second case we note that the D-term can be written as a total derivative
\begin{equation}
\label{Dtermchange}
 \tilde{D}=\frac{1}{2} \frac{\partial}{\partial \eta} \phi_i e^{2 \mu_i \eta}\phi^i
\end{equation}

The function $\phi_i e^{2 \mu_i \eta}\phi^i$ goes to $+\infty$ as $\eta \rightarrow \pm \infty$ and so it has a minimum where the it's derivative must vanish. In this minimum the D-term vanishes. As known, this is why in $U(1)$ theories without an F.I. term or non-Abelian theories, pure D-term SUSY breaking is not possible.

We now look at models where the F-terms are not gauge invariant. In this case the vacuum manifold does not have the enhanced symmetry, and the previous argument does not hold. However, the vacuum manifold contains at least the direction parameterized by $\phi^i=(\phi^i)^{(0)}+z W^{i}$.

Since not all F-terms are gauge invariant, the heavy vector superfield will not be SUSY; for example there will be mixings between gauginos and chiral fermion mass matrices.

One can still investigate how different matrix elements are constrained by gauge invariance:
\begin{equation}
\label{gaugeinvariance}
 W_i(D^a)^{,i}\equiv W_i (T^a)^i_j \phi^j \equiv0
\end{equation}
Because the superpotential must be gauge invariant, note that this is an identity and is not valid only at the minimum of $V_F$. Because of this we can get further identities from differentiating the previous expression:
\begin{equation}
\begin{array}{c}
 W_{ij}(D^a)^{,i}+ W_i(D^a)^{,i}_{j}\equiv0\\
 W_{ijk}(D^a)^{,i}+ W_{ij}(D^a)^{,i}_{k}+W_{ik}(D^a)^{,i}_{j}\equiv0\\
\end{array}
\end{equation}

At the minimum of the $V_F$, the previous expressions implies that:
\begin{equation}
\label{minVfconstraint}
 W_i(T^a)^i_k W^k\equiv- W_{ik}W^k(D^a)^{,i}=0
\end{equation}

We will now see that the D-terms are constant along the pseudo-moduli direction $\phi_i=\phi^{(0)}_i+z W_i$:

\begin{equation}
 D^{a}(z)=D^{\alpha}(0)+2Re(z(\phi^{(0)}_i)(T^{a})^i_jW^j)+|z|^2 W_i (T^{a})^i_jW_j=D^{\alpha}(0)
\end{equation}

The linear term in z vanishes because of( \ref{gaugeinvariance}), and the quadratic term in z vanishes by (\ref{minVfconstraint}).

We will now show that even though the D-terms are constant along that particular pseudo-moduli direction, they can be made to vanish arbitrarily close to it. We will give first a specific example and discuss why this happens and then give a general argument.

Let us take a model considered in \cite{Dine:2006xt}: we have two chiral/antichiral fields  that transform under a U(1), and a singlet. The superpotential is:

\begin{equation}
 W=hS (\phi \tilde{\chi} - \mu^2)+m_1 \phi \tilde{\phi}+m_2 \chi \tilde{\chi}
\end{equation}

Except for S, the untilded fields have charge +1 under the U(1) and the tilded fields have charge -1. The model has an R-symmetry with charges $R(S)=2$, $R(\phi)=R(\tilde{\chi})=0$,$R(\tilde{\phi})=R(\chi)=2$, and the superpotential is general. 

The F-term potential is given by

\begin{equation}
 V_F=h^2 |\phi \tilde{\chi} -\mu^2|^2+m_1^2 |\tilde{\chi}|^2+m_2^2|\phi|^2+|hS \tilde{\chi}+m_1\tilde{\phi}|^2+|hS \phi+m_2\chi|^2
\end{equation}

The model breaks SUSY at tree-level, and the minimum is:
\begin{equation}
 \begin{array}{cc}
V_F=h^2\mu^4 & \text{if} \quad \mu^2<\frac{m_1m_2}{2 h^2}\\
V_F=2 m_1 m_2 \mu^2-\frac{m_1^2 m_2^2}{h^2} & \text{if}\quad \mu^2>\frac{m_1m_2}{2 h^2}
 \end{array}
\end{equation}
This as long as both $m_1$ and $m_2$ are positive. This model has pure F-term SUSY breaking:
In the first case no charged field gets a vev. The $U(1)$ is unbroken and the D-terms vanish.

In the second case the $U(1)$ is broken. The vevs are:
\begin{equation}
 \begin{array}{c}
 \phi=-\frac{\sqrt{m_2}\sqrt{m_1(h^2 \mu^2-m_1 m_2)}}{hm_1}\\
\tilde{\phi}=\frac{\sqrt{m_1(h^2 \mu^2-m_1 m_2)}S}{m_1 \sqrt{m_2}}\\
\chi=\frac{\sqrt{m_1(h^2 \mu^2-m_1 m_2)}S}{m_1 \sqrt{m_2}}\\
\tilde{\chi}=-\frac{\sqrt{m_1(h^2 \mu^2-m_1 m_2)}}{h\sqrt{m_2}}\\
\end{array}
\end{equation}
for any value of S.

The flat direction is parameterized by:
\begin{equation}
 \begin{array}{c}
S=S^{(0)}-z\frac{m_1 m_2}{h}\\
\tilde{\phi}=\tilde{\phi}^{(0)}-\frac{\sqrt{m_2}\sqrt{m_1(h^2 \mu^2-m_1 m_2)}}{h}z\\
\chi=\chi^{(0)}-\frac{\sqrt{m_2}\sqrt{m_1(h^2 \mu^2-m_1 m_2)}}{h}z\\
\end{array}
\end{equation}

If we replace these vevs into $V_D$, we get that $V_D=\frac{1}{2}g^2\frac{(m_1^2-m_2^2)^2(m_1m_2-h^2\mu^2)^2}{h^4 m_1^2 m_2^2}$, which is generally non-zero unless $m_1=m_2$. Note that this is a constant and independent of z, as we have seen it should be.

We will now deform the vevs by $\epsilon$ in the following way:
\begin{equation}
 \begin{array}{c}
 ×\phi=-\frac{\sqrt{m_2}\sqrt{m_1(h^2 \mu^2-m_1 m_2)}}{hm_1}\\
×\tilde{\phi}=\frac{\sqrt{m_1(h^2 \mu^2-m_1 m_2)}S}{m_1 \sqrt{m_2}}-\epsilon\\
×\chi=\frac{\sqrt{m_1(h^2 \mu^2-m_1 m_2)}S}{m_1 \sqrt{m_2}}+\epsilon\\
×\tilde{\chi}=-\frac{\sqrt{m_1(h^2 \mu^2-m_1 m_2)}}{h\sqrt{m_2}}\\
S=\frac{(m_1^2-m_2^2)\sqrt{m_1(h^2 \mu^2-m_1 m_2)}}{4 m_1 \sqrt{m_2} h^2\epsilon}\\
\end{array}
\end{equation}

In this vacuum 
\begin{align}
 &V_F=2m_1 m_2 \mu^2-\frac{m_1^2 m_2^2}{h^2}+(m_1^2+m_2^2) \epsilon^2\\
&V_D=0
\end{align}

So, if $\epsilon\rightarrow 0$, we solve $V_D=0$ and minimize $V_F$.

 More generically will show that one can find a small deformation $\epsilon$ of the pseudo-moduli space, where the change in the D-terms is proportional to $\epsilon z$, while the change in the F-terms will only be proportional to $\epsilon$. One can then take $\epsilon \rightarrow 0$ with  $z \epsilon$ fixed, to solve $V_D=0$, and $V_F$ will approach it's minimum value. This means that if we ignore D-terms the model breaks SUSY at tree-level through F-terms, but as soon as we include D-terms they either vanish identically on the pseudo-moduli space or there is a runaway to broken SUSY. 

The stabilization of this runaway direction has to be checked because in these models one has the extra contribution from gauge fields to the Coleman-Weinberg potential, and these are not positive definite. This fact can be seen, for example, in \cite{Intriligator:2007py} and in the last example model we show. Some cases where some metastable vacua with both F and D-terms 
and no flat directions have been built \cite{Dienes:2008rg}.

We will consider that a gauge rotation of the fields has been made in such a way that there is only one non-vanishing D-term in a direction defined by an element of the Cartan subalgebra of the group, very much like when we dealed with the case that all the F-terms were gauge invariant. 

We will chose that the deformation is a complexified gauge transformation in the direction defined by the same element of the Cartan subalgebra. This transformation will generically change both $V_F$ and $V_D$.

Let T be the element of the Cartan subalgebra, then $T=diag(\{ \mu_i\})$.
Under this complexified gauge transformation The F-terms change as
\begin{equation}
 \tilde{W}_i(z)=W_j(z) (e^{- \alpha_b T^b})^j_i=(W(0)^{\dagger}.(e^{- \alpha T})_i)=(W^{\dagger}.diag(\{ e^{\alpha \mu_i}\}))_i
\end{equation}

Where, again, we have used $\phi$ and dot product as a notation for the sum over all the fields and the generator is in the appropriate representation.

Then, the change in the potential is
\begin{equation}
 V_F(\alpha)=W^{\dagger}.diag(\{ e^{\alpha \mu_i}\}).W \approx V_F(0)+\alpha^2 W^{\dagger}.diag(\{ {\mu_i^2}\}).W+O(\alpha^3)
\end{equation}
This is because, as we've seen, the terms linear in the generator of the gauge group vanish. The previous example where we had a $U(1)$ gauge group agrees with this expression.

We now use the expression (\ref{Dtermchange}) evaluated at $(\phi^i)'=\phi^i+zW^i$.

We get that:
\begin{align}
 \tilde{D}(z)&=\phi^{\dagger}.diag(\{\mu_i\}).\phi\\
&-2Im(\alpha) (\phi^{\dagger}.diag(\{\mu_i^2\}).\phi+2 Re(z\phi^{\dagger}.diag(\{\mu_i^2\}).W)+|z|^2 W^{\dagger}.diag(\{\mu_i^2\}).W)
\end{align}
When z is very large this is solved by:

\begin{equation}
 Im(\alpha)=\frac{\phi^{\dagger}.diag(\{\mu_i\}).\phi}{2 |z|^2 W^{\dagger}.diag(\{\mu_i^2\}).W} 
\end{equation}

The solution $\alpha$ is small and the corrections in $\alpha^2$ we ignored are negligible if we take $z \rightarrow \infty$. Note that $W_i \mu_i^2 W^i=\sum_i|\mu_i W^i|^2$ and should be greater than zero if SUSY is broken and the F-terms are not gauge invariant.

The conclusion is that in this class of models it's not possible to have a model where the global minimum has combined F and D-term SUSY breaking.

One should, of course, be careful to check if the description being used is correct and one doesn't go to a strong coupling regime or if one is using an effective theory, go outside its regime of validity.

\section{Adding non polynomial terms to the superpotential}

In the last case, the discussion relied in the existence of the flat direction, so one question is: what happens if we do not have a flat direction. One such possibility is to consider the present model, but instead of a U(1) gauge symmetry, one considers a $SU(N_c)$ gauge symmetry with $N_c>2$. In this case we have a SQCD theory with 2 flavours and a singlet and $N_c>N_F$. This theory has a dynamically generated term in the superpotential that comes from instanton contributions, and is not a polynomial of degree 3 (for simplicity we take $N_c=3$).

The full superpotential is:

\begin{equation}
 W=hS (\phi \tilde{\chi} - \mu^2)+m_1 \phi \tilde{\phi}+m_2 \chi \tilde{\chi}+\frac{\Lambda^7}{\phi.\tilde{\phi} \chi. \tilde{\chi}-\chi.\tilde{\phi} \phi. \tilde{\chi}}
\end{equation}

At tree-level the superpotential is general and has an R-symmetry (this disallows shifts in the vev of S to absorb the linear term). However, for $N_f<N_C$ this symmetry is anomalous because of the dynamically generated ADS contribution, and the R-symmetry is broken to a discrete subgroup.

One may wonder this model breaks SUSY globally. If this were to happen (SUSY is unbroken), we could integrate out the gauge degrees of freedom and use the low energy effective description with mesons:

\begin{equation}
 W=hS (M_{12} - \mu^2)+m_1 M_{11}+m_2 M_{22}+\frac{\Lambda^7}{det(M)}
\end{equation}
The F-term equations are:

\begin{equation}
\begin{array}{c}
F_S=h(M_{12}-\mu^2)\\
F_{M_{11}}=m_1-\frac{\Lambda^7 M_{22}}{det(M)^2}\\
F_{M_{12}}=h S+\frac{\Lambda^7 M_{21}}{det(M)^2}\\
F_{M_{21}}=\frac{\Lambda^7 M_{12}}{det(M)^2}\\
F_{M_{11}}=m_2-\frac{\Lambda^7 M_{11}}{det(M)^2}\\ 
\end{array}
\end{equation}

These equations can be solved asymptotically with

\begin{equation}
 \begin{array}{c}
 S=-\frac{\Lambda^7 M_{21}}{h det(M)^2}\\
M_{11}=\frac{m_2 \Delta^8}{\epsilon^2 \Lambda^7}\\
M_{12}=\mu^2\\
M_{21}=\frac{\Delta^4+\frac{m_1 m_2\Delta^{12}}{\epsilon^3\Lambda^{16}}}{\epsilon \mu^2}\\
M_{22}=\frac{m_1 \Delta^8}{\epsilon^2 \Lambda^7}\\
\end{array}
\end{equation}

where $\Delta$ is some finite mass scale and $\epsilon\rightarrow0$ is some dimensionless parameter. Note that in this limit all vevs but $M_{12}$ go to infinity.

In the D-flat directions the Kahler potential is given by:
\begin{equation}
K=\sqrt{M^{\dagger}M}
\end{equation}
Where this is valid as long as the theory is weakly coupled (i.e) mesonic vevs are larger than $\Lambda$.

The second derivative of the Kahler potential goes to 0 as $\epsilon\rightarrow0$. One can, for simplicity take $m_1=m_2=m$, diagonalize M with unitary transformations and do the computation explicitly (using Mathematica for example) and see that in the runaway direction the $V_F=K^{-1}_{ab}W^a W^b$ goes to $2 m^2 \mu^2$.

This can be understood since as $\epsilon$ becomes very small, the scale at which the gauge symmetry is broken becomes large and the theory is weakly coupled. The reason why the potential doesn't slope to 0 for very large vevs of the meson field is easy to understand since in this regime the theory should be weakly coupled and the microscopic description should be good. Furthermore, in this regime, the instanton contribution is rather small compared with the tree-level superpotential, and there is an approximate R-symmetry that guarantees that SUSY is broken. 

The reason why we can't just expand the Kahler potential around some scale to get a ``canonical'' Kahler potential for the meson fields (normalized by the square of the scale we are expanding around), is that there is no scale around which we can do this, i.e. there is no scale around which the minimization of $V_F$ will justify the fact that we've neglected higher order terms in the expansion of the Kahler potential ($V_F$ will be a function that slopes to 0 when the fields go to infinity). 

\subsubsection*{Witten Index}

One may wonder if the statement that this model breaks SUSY is in disagreement with the Witten index argument, but note that taking $m_1$ and $m_2$ to infinity (the mass matrix has maximal rank) allows us to integrate out the quark fields ($m_{susy}\ll m_1,m_2$). The scalar field is then light and survives to low energy, doing this integration one sees that at low energies the effective superpotential is $W=-hS\mu^2+N_C\Lambda_{SYM}^3$. In this case the $Det(\mathbb{M})=m_1 m_2$, where $\mathbb{M}$ is the quark mass matrix (as in \cite{RevModPhys.72.25}), and $\Lambda_{SYM}$ does not depend on the vev of S. There is no mass term for S in the superpotential and because of the linear term, the model breaks SUSY. So the Witten index must vanish.

\subsubsection*{Numerical Calculations}

Using Mathematica one can compute the eigenvalues of the matrix $M^{\dagger}M$ and do the square root. The sign is determined by the physical requirement that the potential should be bounded from bellow. It's straightforward to compute the Kahler metric and invert it to compute the $V_F$. 
One can then give values to the parameters and Mathematica provides different numerical algorithms that help finding the minimum of the potential.

As a test example, we use $m_1=1.2$,$m_2=0.45$,$h=0.8$,$\Lambda=0.05$ in units of $\mu$, and $g=0.75$. Using the mesonic degrees of freedom, we find that there is a minimum for:

\begin{equation}
\begin{array}{c}
S=-23.36\\
M_{11}=2.43\\
M_{12}=0.16\\
M_{21}=101.10\\
M_{22}=6.49\\
\end{array}
\end{equation}

 with $V_F= 0.62\mu^4$. However, this minimizes $V_F$ along the directions where $V_D=0$, and may not be the right procedure. Instead of improving our understanding of the Kahler potential around the classical D-flat ``pseudo-moduli space'', as in \cite{Seiberg:2008qj,Elvang:2009gk}, we follow directly to the microscopic description.

We now minimize the potential using the microscopic description, and allowing for non-vanishing D-terms.
The $V_F$ potential is:
\begin{align}
 V_F&=h^2|\phi.\tilde{\chi}-\mu^2|^2+|m_1 \phi_i-\frac{\Lambda^7 (\chi.\tilde{\chi} \phi_i- \phi.\tilde{\chi} \chi_i)}{(\phi.\tilde{\phi} \chi. \tilde{\chi}-\chi.\tilde{\phi} \phi. \tilde{\chi})^2}|^2 +|h S \tilde{\chi}_i+m_1\tilde{\phi}_i-\frac{\Lambda^7 (\chi.\tilde{\chi} \tilde{\phi}_i -\chi.\tilde{\phi} \tilde{\chi}_i)}{(\phi.\tilde{\phi} \chi. \tilde{\chi}-\chi.\tilde{\phi} \phi. \tilde{\chi})^2}|^2\nonumber\\
&+|h S \phi_i+m_2\chi_i-\frac{\Lambda^7 (\phi.\tilde{\phi} \chi_i -\chi.\tilde{\phi} \phi_i)}{(\phi.\tilde{\phi} \chi. \tilde{\chi}-\chi.\tilde{\phi} \phi. \tilde{\chi})^2}|^2+|m_2\tilde{\chi}_i-\frac{\Lambda^7 (\phi.\tilde{\phi} \tilde{\chi}_i -\phi.\tilde{\chi} \tilde{\phi}_i)}{(\phi.\tilde{\phi} \chi. \tilde{\chi}-\chi.\tilde{\phi} \phi. \tilde{\chi})^2}|^2
\end{align}

Where a sum over i is assumed. We chose some gauge fixing conditions: $\phi_2=0$, $\phi_3=0$, $\chi_3=0$ and minimized the potential with respect to the other fields, where the lower case index is a color index. Numerically we found that $\tilde{\phi_3}=0$,$\tilde{\chi}_3=0$. 

The non-vanishing vevs are given (in units of $\mu$) by:
\begin{equation}
\begin{array}{c}
S=-1.51\times 10^{-3}\\
\phi_1=0.50\\
\tilde{\phi}_1=1.4\times 10^{-3}\\
\tilde{\phi}_2=1.36\times 10^{-4}\\
\chi_1=1.66\times 10^{-3}\\
\chi_2=-2.28\times 10^{-3}\\
\tilde{\chi}_1=0.35\\
\tilde{\chi}_2=-0.67\\
\end{array}
\end{equation}

And $V_F+V_D=0.58\mu^4$, which is lower than the previous result. This vacuum has $V_F=0.54\mu^4$ and $V_D=0.04\mu^4$, and so both D and F-terms are non-zero. This means that by moving away from the ``D-flat'' directions we were able to find a global minimum where SUSY is broken, and because the scale at which the gauge group is broken is much larger than $\Lambda$, the theory is weakly coupled.

We note that the scale at wich gauge symmetry is broken is only one order of magnitude above the strong coupling regime, so quantum corrections are not negligible. One can change the parameters (for example, decreasing g will increase the vevs of the quarks), however, this does not change the fact that SUSY is broken. As a consequence of this fact, this model should be taken in the sense of an existence proof more than anything else.

To summarize what we have done: we have took a model that had an R-symmetry and wrote the most general superpotential consistent with this. We noticed that for $N_f<N_c$ there are non-perturbative corrections to the potential that in practice make this symmetry anomalous. Despite having no continuous R-symmetry the model breaks SUSY because the superpotential is not generic and, for some range of the parameters, the global vacuum of the theory has a combined F and D-term SUSY breaking vacuum.

The way this work makes contact with the work of  \cite{Izawa:1996pk,Intriligator:1996pu} is that we can look at the gauge singlet as a Lagrange multiplier that deforms the ``classical moduli space'' of SQCD. On the other hand, the singlet also changes the behavior of the potential at infinity allowing the Witten index to change from the SQCD case. One can readily see that in this model it has to vanish (by looking at the case where the quark masses are much larger than the SUSY breaking scale).

A detailed study of the phenomenology and inclusion of messengers in this model is postponed for later. Some recent work in semi-direct mediation was done in \cite{Ibe:2009bh} using very similar models to the one we presented here, except that they use it for the messenger sector. In this work they were able to retrofit the model to generate gaugino masses at leading order in $F/M$.
A different approach to mediating the SUSY breaking effects to the MSSM was done in \cite{Raby:1998bg} (the messengers coupling to the hidden sector is an irrelevant operator suppressed by a high mass).

\section{Getting D-terms at loop level, the runaway case}

\subsection{The model}

\paragraph{}
We will follow refs.\cite{Abel:2007jx,Abel:2007nr,Abel:2008gv}. In the magnetic description of the theory, the baryon deformation consists in having an ISS model with magnetic gauge group $N=2$ and $N_{f}=7$, where we've added to the superpotential a quadratic term in the quark field:

\begin{equation}
W=h( \Phi_{ij}\varphi_i.\tilde{\varphi_{j}}-\mu^{2}_{ij}\Phi_{ji})+m\epsilon_{ij} \epsilon^{ab} \varphi_a^i \varphi_b^j 
\end{equation}

The last term is a baryon of $SU(2)$ and is a relevant deformation of the low energy magnetic description. The quark fields have a charge that is not 0 or 2, allowing for spontaneous R-symmetry breaking. 

The masses $\mu_{ij}$ break flavour symmetry from $SU(7)$ down to $SU(2) \times SU(5)$, and can be written as: 

\begin{equation}
\mu^2_{ij}=\left(
\begin{tabular}{cc}
$\mu^2 \mathbb{I}_2$ & $0$ \\
$0$ & $\hat{\mu}^2 \mathbb{I}_5$
\end{tabular} \right)
\end{equation}
with $\mu^2 >\hat{\mu}^2$ to keep the $SU(5)$ symmetry unbroken.
We can write the fields as:

\begin{table}[h]
\begin{center}
\begin{tabular}{|c | c | c | c|}
\hline
  & $SU(2)_f$ & $SU(5)_f$ &$U(1)_R$ \\ \hline \hline
 $
 S \equiv \left(
  \begin{array}{cc}
   Y & Z  \\
   \tilde{Z} & X \\
  \end{array} \right)
$ &
 $
  \left(
  \begin{array}{cc}
   Adj+1  & \overline{\square}  \\
   \square & 1\\
  \end{array} \right)
$ & $
  \left(
  \begin{array}{cc}
   1 & \overline{\square}  \\
   \square & Adj+1 \\
  \end{array} \right)
$ &2 \\

$
 \varphi \equiv \binom{\phi}{\rho}
$ & $
  \binom{\square}{1}
$ &   $\binom{1}{\square}
$&$\frac{2}{3}$ \\

$
 \tilde{\varphi} \equiv \binom{\tilde{\phi}}{\tilde{\rho}}
$ & $
  \binom{\overline{\square}}{1}
$ &$
  \binom{1}{\overline{\square}}
$ & $-\frac{2}{3}$ \\
\hline
\end{tabular}
\end{center}
\end{table}

Classically there's a runaway direction parameterized by:

\begin{equation}
\begin{array}{ll}
\langle \varphi \rangle= \frac{\mu^2}{\xi}\mathbb{I}_2, & \langle\tilde{\varphi}\rangle=\xi\mathbb{I}_2\\
\langle\eta\rangle=\frac{-2m}{h(\left(\frac{\xi}{\mu}\right)^2+(\left(\frac{\mu}{\xi}\right)^2)}\mathbb{I}_2, &\langle X \rangle= \chi\mathbb{I}_5 \\
\end{array}
\end{equation}

when $\xi$ goes to $\infty$.

Motivated by these constrains, we will minimize the potential along the directions:

\begin{equation}
\begin{array}{cc}
\phi=\kappa \mathbb{I}_2, & \tilde{\phi}=\xi \mathbb{I}_2\\
Y=\eta\mathbb{I}_2, & X= \chi \mathbb{I}_5
\end{array}
\end{equation}

And check that the pseudo-moduli have positive masses in the minimum we find.

%
The mechanism through which SUSY breaking is communicated to the SSM is direct gauge mediation:
we gauge the unbroken $SU(5)$ of flavour and the fields charged under this symmetry will act as messengers. These will be of two types: fundamentals($\rho$,$\tilde{\rho}$,$Z$ and $\tilde{Z}$) and the adjoint $X$. The fundamentals will generate masses for the gauginos at one loop and masses for the scalars at two loops in SSM, while the adjoints will contribute to gaugino masses through an effective one loop diagram \cite{Abel:2008gv}. For this particular model the contribution of the adjoints to gaugino masses is negligible when compared to the contribution of the fundamentals (even when we take into account the effects of D-terms and their derivatives). (It was recently noted that in the case where the gauge group is $SO(N)$ there is no runaway and this deformation probably does not play a role in getting bigger gaugino masses). For simplicity, in the results presented, we will only show the contribution of the fundamentals.

We define reduced masses $m_{1/2}$ and $m_0$ for the gauginos and the scalars:
\begin{equation}
\begin{array}{c}
 m_{\lambda_A}(\mu):=\frac{g^2_A}{16\pi^2}m_{1/2}\\
m_{\bar{f}}(\mu):= \sum_A \frac{g^4_A}{(16\pi^2)^2}C_A S_A m_0^2\\
\end{array}
\end{equation}

Where $C_A$ and $S_A$ are the Casimir and Dynkin index of the gauge group. And $m_{\lambda_A}(\mu)$ and $m_{\bar{f}}(\mu)$ can be computed from the usual diagrams.

\subsection{The computation of the effective potential}
\paragraph{}
The Coleman-Weinberg potential is given by the following expression:

\begin{align}
V_{eff}^{(1)}= &\frac{1}{64\pi^2}Str(\mathcal(M)^4log(\frac{\mathcal{M}^2}{\Lambda_{UV}^2})\nonumber\\
\equiv& \frac{1}{64\pi^2}( Tr(m_{sc}^4 log(\frac{m_{sc}^2}{\Lambda_{UV}^2})-2\, Tr(m_{f}^4 log(\frac{m_{f}^2}{\Lambda_{UV}^2})+3\, Tr(m_{v}^4 log(\frac{m_{v}^2}{\Lambda_{UV}^2}))
\end{align}

The mass matrices are given by:

\begin{equation}
M_0^2=\left(
\begin{array}{cc}
W^{ab} W_{bc} + D^{\alpha a} D^{\alpha}_c + D^{\alpha a}_c D^{\alpha} & W^{abc} W_b + D^{\alpha a} D^{\alpha c} \\
W_{abc}W^b+D^{\alpha}_a D^{\alpha}_c & W_{ab}W^{bc}+D^{\alpha}_a D^{\alpha c}+D^{\alpha c}_a D^{\alpha}
\end{array}
\right);
\end{equation}

\begin{equation}
\begin{array}{cc}
M_{1/2}^2=\left(
\begin{array}{cc}
W^{ab}W_{bc}+2 D^{\alpha a}D^{\alpha}_c & -\sqrt(2) W^{ab} D^{\beta}_b\\
-\sqrt(2) D^{alfa b}W_{bc} & 2 D^{\alpha c} D^{\alpha}_c
\end{array}
\right), &

M_1^2=D^{\alpha }_a D^{\beta a}+D^{\alpha a}D^{\beta}_a;\\
\end{array}
\end{equation}
 where $W_c \equiv \frac{\partial W}{\partial \Phi^c}$ is the derivative of the superpotential with respect to the scalar component of the superfield $\Phi^c$, $D^{\alpha}=g z_aT^{\alpha a}_b z^b$ and $D^{\alpha}_a$ is the derivative of the D-term with respect to $\Phi^a$. And, as before, indices are raised and lowered by complex conjugation. At one loop the effective potential is $V_{eff}=V_0+V^{(1)}_{cw}$.

The difference of the computations presented below is the inclusion of D-term and their derivatives in the calculation of the Coleman-Weinberg potential. In other words, we take into account not only the dependence of the effective potential on the superpotential parameters but also the ISS gauge coupling.

As we've said before, the reason why this is important is that the masses of the particles in the Higgsed vector multiplet will not be supersymmetric (even in the absence of D-terms),since they couple to some F-terms that are not gauge singlets. This effect will then feed into the Coleman-Weinberg and the minimization of the potential. The parameter that controls the relative size of these contributions to the ones coming from chiral fields (that were not Higgsed) is $k\equiv \left(\frac{g}{h}\right)^2$. This is because contributions to the Coleman-Weinberg scale with $F\times m^2$, where F are the F-terms that break SUSY, and m is the mass of the particle; for chiral fields this is $h^4$ and for fields in the vector multiplet this is $h^2 g^2$, and the ratio is the parameter k defined before. If k is large (but both h and g are small), contributions coming from the Coleman-Weinberg could change the tree level constraints by a sufficient amount to allow the generation of larger gaugino masses.

When computing the one loop correction one needs to know the masses of all particles. In this particular model, because we have a runaway direction, this presents a problem. We have some global symmetries that are spontaneously broken. For each of the broken symmetry generators there should be a goldstone boson. However since we are not at the tree level minimum of the potential, these modes will not be massless. This only happens after the stabilization of the potential by quantum corrections.
These will shift the mass of the modes corresponding to broken global symmetry generators by a small amount. What makes this relevant for these modes is that since this shift can be positive, their tree level masses can be negative. This is just a reflection that only the full quantum corrected potential has a well defined vacuum. In view of this we will not try to ensure the positivity of the masses for these modes (which we will call Goldstone modes anyway), and estimate the error by looking at the minimum of the following functions:

\begin{equation*}
\begin{array}{l}
V_{cw}=V_{cw}(Massive\, modes)\\
V^{\pm}_{cw}=V_{cw}(Massive\, modes) \pm V_{cw}(Goldstone\, modes)
\end{array}
\end{equation*}
This is similar to including the one-loop corrected mass for Goldstone modes and tree-level masses for all other fields when computing the Coleman-Weinberg potential.

By looking at several cases with different parameter choices, we've seen that usually the error introduced in this approximation is small, and probably of the same order of two loop corrections. We show two representative cases (the diagonalization of the mass matrices was done numerically and the error is omitted whenever it's smaller than numerical sensitivity):

\begin{table}[h]
\begin{tabular}{|c|c|c|c|c|}
\hline
$k$ & $\eta$ & $\kappa$ &  $\xi$  & $\chi$ \\
\hline
\hline
$0$  &  $-0.1601$ & $0.5674 \pm 1\times 10^{-7}$ & $7.0403 \pm 3\times 10^{-6}$ & $-0.3157 \pm 4\times 10^{-7}$\\
\hline
$0.36$ &  $-0.1602$ & $0.5692$  & $7.0570 $& $-0.3177$\\
\hline
$1.0 $&$ -0.1566 \pm 2\times 10^{-7}  $&$ 0.5671\pm 1\times 10^{-7}$  & $7.1914\pm 1\times 10^{-6}$ & $-0.3155\pm 5\times 10^{-6}$\\
\hline
\end{tabular}
\caption{Values of the vev's for the baryon deformation using $h=1$, $m=1$,$\mu=2$ and $\hat{\mu}=1$, $k=\left(\frac{g}{h}\right)^2$.}
\end{table}

\subsection{Results}

\paragraph{} We now study what happens when we switch on the gauge coupling:

As one could expect, and can see in tables 1-3 the position at which the vacuum is stabilized depends on the gauge coupling. The violation of the classical constraints increase as we increase k (there is a decrease at the beginning because h and g contributions to the violation of the tree level constraints have opposite signs). There is, however a maximum value for k, after which the potential is not stabilized at one loop, for the point that we show this is $k \approx 4$ . The reason why this happens is that the F-terms that exist along the runaway cause a mixing in the fermionic mass matrix between the $N_c^2-1$ gauginos of the hidden sector and the fermions. This is proportional to $g^2 h^2$ and comes with a factor of -2 in the Coleman-Weinberg. In the scalar mass matrix, there is a mixing of the $N_c^2-1$ ``Higgsed'' particles proportional to $g^2 h^2$ and this comes with a factor of 1. There is also the usual mixing due to F-terms and this is proportional to $h^4$. So, there are directions in field space where the Coleman-Weinberg behaves as $h^4(1-\alpha k)$. If $\alpha$ is large enough the potential will not be stabilized.

\begin{table}[h]
\begin{center}
\begin{tabular}{|c|c|c|c|c|}
\hline
$k$  & $\eta$ & $\kappa $&  $\xi $ &$ \chi $\\
\hline
\hline
$0   $ & $-0.1634$ & $0.3617$ & $11.0533$ & $-0.3258$\\
\hline
$0.44$ & $-0.1635$ & $0.3622$ & $11.0660$ & $-0.3268$\\
\hline
$1.78$  & $-0.1578$ & $0.3581$ & $11.3329$ & $-0.3194$\\
\hline
$3.36$  & $-0.1259$ & $0.3229$ & $12.8147$ & $-0.2600$\\
\hline

\end{tabular}
\end{center}
\caption{Values of the vev's for the baryon deformation using $h=0.6$, $m=1.5$, $\mu=2$ and $\hat{\mu}=1$.}
\end{table}

\begin{table}[h]
\begin{center}
\begin{tabular}{|c|c|c|c|}
\hline
$k$  & $\eta_{c}-\eta$ & $\kappa_{c}-\kappa$ &  $m_{\frac{1}{2}}$\\
\hline
\hline
$0 $  & $3.2\times 10^{-5}$ & $1.69\times 10^{-4}$  & $3.85\times 10^{-4}$\\
\hline
$0.44$ & $4.55\times 10^{-4} $ &$-7.7\times 10^{-4}$ &$ 4.02\times 10^{-4}$\\
\hline
$1.78$ & $2.36\times 10^{-3}$ &$-5.16\times 10^{-3}$  &$ 5.29\times 10^{-4}$\\
\hline
$3.36$ & $4.23\times 10^{-3}$ &$-1.08\times 10^{-2}$  &$ 8.00\times 10^{-4}$\\
\hline
\end{tabular}
\end{center}
\caption{Evolution of the violation of constraints with k and soft gaugino masses.}
\end{table}

We now turn to the computation of the ratio of gaugino vs scalar masses. Before going to the numerical calculations we can understand what will happen by noting that in leading order of F/M, gaugino masses vanish when the classical constraints are satisfied. So we expect that they increase when the violations of these constraints increase. By looking at table 3, we see that we expect a decrease of the ratio of gaugino vs scalar mass for small k and then an increase. This is indeed what happens:

\begin{table}[h]
\begin{center}
\begin{tabular}{|c|c|c|}
\hline
$k$   & $m_0$ & $m_{\frac{1}{2}}$/$m_0$\\
\hline
\hline
$0$   & $0.402$ & $9.58\times 10^{-4}$\\
\hline
$0.44$ & $0.402$ & $1.00\times 10^{-3}$\\
\hline
$1.78$ & $0.394$ & $1.34\times 10^{-3}$\\
\hline
$3.36$ & $0.356$ & $2.23\times 10^{-3}$\\
\hline
\end{tabular}
\end{center}
\caption{Change of the ratio of gaugino to scalar reduced masses with k, $h=0.6$, $m=1.5$, $\mu=2$ and $\hat{\mu}=1$}
\end{table}

\begin{figure}
\centering
\includegraphics[angle=0]{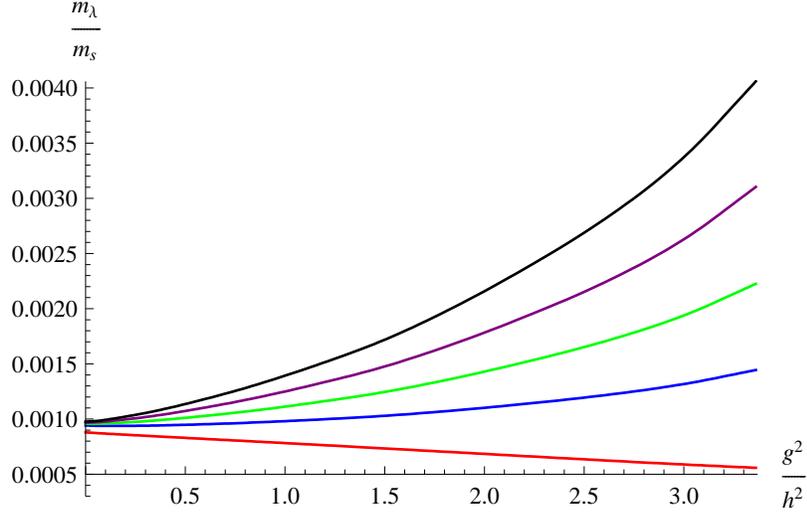}
\caption{Impact of the gauge coupling on the ratio $\frac{m_{\lambda}}{m_s}$ where $m_{\lambda}$ is the reduced gaugino mass and $m_{s}$ is the reduced scalar mass, for $m=0.3$ in red, $m=1$ in blue, $m=1.5$ in green, $m=2$ in purple and $m=2.5$ in black.}
\end{figure}

One could hope to reduce the hierarchy qualitatively between gaugino and scalar masses when $k>1$, but even though there is a decrease it is not enough: unfortunately, in the promising regime of large k (but both h and g are small), we stop being able to stabilize the potential with one loop corrections.

In these cases, even though there are effects coming from mixing derivatives of D-terms and F-terms, there are no non-vanishing D-terms because of an unbroken $SU(2)_v$. Under this global symmetry the quarks transforms as fundamental $\times$ anti-fundamental, i.e. the only vev it can get is for the singlet component, and this must be proportional to the identity matrix. This symmetry could have been spontaneously broken but, numerically, we didn't find this to happen. 

We generate non-vanishing D-terms by breaking the $SU(2)_f$ global symmetry by mass terms, which also breaks the $SU(2)_v$. These contribute to the tree level potential, so one might thing that they will not be generated. The reason why they do appear, is because there is a decrease in the Coleman-Weinberg that compensates for this increase.
The mass matrix is now:
\begin{equation}
\mu^2_{ij}=\left(
\begin{tabular}{ccc}
$\mu_1^2 $ & $0$ & $0$ \\
$0$  &  $\mu_2^2 $ & $0$ \\
$0$  & $0$ & $\hat{\mu}^2 \mathbb{I}_5$
\end{tabular} \right)
\end{equation}

The vev is parameterized as before except that now $\phi$, $\tilde{\phi}$ and Y are $2\times2$ matrices with different diagonal values: 
\begin{equation}
\begin{array}{cc}
\phi=\left(
\begin{tabular}{cc}
$\kappa_1 $ & $0$  \\
$0$  &  $\kappa_2 $ \\
\end{tabular} \right)
, & \tilde{\phi}=\left(
\begin{tabular}{cc}
$\xi_1 $ & $0$  \\
$0$  &  $\xi_2 $ \\
\end{tabular} \right)
\\
Y=\left(
\begin{tabular}{cc}
$\eta_1 $ & $0$  \\
$0$  &  $\eta_2 $ \\
\end{tabular} \right)
, & X= \chi \mathbb{I}_5
\end{array}
\end{equation}

We have checked that non-diagonal vev's for $\phi$, $\tilde{\phi}$ and Y are not generated after the inclusion of the Coleman-Weinberg potential, hence we used this simpler form for the minimization.
 
With this ansatz for the vacuum only the third component of the D-terms $D^3$ can be non-zero. Also here we don't see a qualitative difference and gaugino masses remain small.

\begin{table}[h]
\begin{center}
\begin{tabular}{|c|c|c|c|c|c|c|c|}
\hline
$k$   & $\kappa_1$ & $\kappa_2$ &  $\xi_1$ & $\xi_2$ &       $\eta_1$         &       $\eta_2$        &        $\chi$ \\
\hline
\hline
$0 $  &  $0.66$    & $0.30$     & $13.63$    &  $13.29$  &  $-1.01\times 10^{-1}$ & $-2.27\times 10^{-1}$ &  $-3.37\times 10^{-1}$ \\
\hline
$0.735$&  $0.67$    & $0.28$     & $14.60$    &  $14.58$  &  $-0.99\times 10^{-1}$ & $-2.24\times 10^{-1}$ &  $-4.71\times 10^{-2}$ \\
\hline
$1.306$&  $0.64$    & $0.28$     & $14.31$    &  $14.29$  &  $-0.91\times 10^{-1}$ & $-2.04\times 10^{-1}$ &  $-4.08\times 10^{-2}$ \\
\hline
\end{tabular}
\end{center}
\caption{Change of the vacuum position with k, for $\mu_1=3.0$, $\mu_2=2.0$, $\hat{\mu}=1$, $m=1.6$, $h=0.7$}
\end{table}

\begin{table}[h]
\begin{center}
\begin{tabular}{|c|c|c|}
\hline
$k$       & $D^3/F$                     & $m_{\frac{1}{2}}$/$m_0$\\
\hline
\hline
$0 $      &  $0.0$                    & $1.20\times 10^{-3}$  \\
\hline
$0.735$   &  $-3.2\times 10^{-3}$    & $1.21\times 10^{-3}$  \\
\hline
$1.306$   &  $-4.1\times 10^{-3}$    & $1.61\times 10^{-3}$  \\
\hline
\end{tabular}
\end{center}
\caption{Change in gaugino mass and D-terms with k. The D-term was rescaled by the SUSY breaking scale.}
\end{table}

\section{Conclusions}

\paragraph{}

The motivation for this work was to try to evade the results of \cite{Komargodski:2009jf}, where they show that when the superpotential is renormalizable and SUSY is broken through F-terms, metastability is unavoidable if one wants to use gauge mediation to generate large gaugino masses in the MSSM.

We investigated the possibility of having models that break SUSY through F and D-terms (very much in the style of \cite{Seiberg:2008qj,Elvang:2009gk}). It was seen that this does not happen if the superpotential is renormalizable and the Kahler potential is canonical. This also will not happen if the Kahler potential is canonical and the non-vanishing F-terms are gauge invariant. (Metastable vacua with non-vanishing F and D-terms are possible, but care needs to be taken regarding flat directions since often these will not be stabilized by one loop corrections.)

We showed that when these condition are not met one can have models that break SUSY in the global vacuum of the effective description with both non-vanishing F and D-terms: SQCD+singlet and baryon deformed ISS. 

In the first case, and by changing the parameters, one goes from a pure F-term SUSY breaking to a combined F and D-term SUSY breaking model. This is not in contradiction with the Witten index argument since the presence of the singlet changes the potential at infinity. In particular, one can choose a region of parameter space where we can integrate out the quarks and at low energies one has a gauge singlet with a classically flat (but non-zero) potential. 

For our particular choice of parameters, gauge symmetry was broken only one order of magnitude above the dynamical scale of the gauge theory, so quantum corrections to the potential are not negligible. This could be improved by changing the parameters (eg. reducing the gauge coupling increases the size of some of the quark vevs). Due to the existence of several tree-level minima with similar energies, study of Coleman-Weinberg corrections is very likelly to be important. This is left for future work.

This model does not include messenger fields however. If one allows for metastability, one can couple a vector-like pair of hidden sector singlets and $SU(5)$ fundamentals to the gauge singlet and have ordinary gauge mediation (see \cite{Giudice:1998bp} for a review). It may be possible to do things in such a way that it makes the discrete R-symmetry\cite{Dine:2009sw} and D-terms present in the model useful. 

This model is another example that combined F and D-term SUSY breaking can be achieved (in a theory with vector-like matter), but more work is required to build a phenomenologically viable model.

The second case is fairly similar to the first, but the vev's of D-terms are loop supressed with respect to the F-terms. One could hope that by increasing the ISS gauge coupling with respect to the ISS parameter h, the departure from the conditions in \cite{Komargodski:2009jf} would be sufficient to generate larger gaugino masses. Unfortunately, in this case, and as in one of the models presented in \cite{Intriligator:2007py}, there is a maximum value of g for which the runaway is stabilized. For this model, similar gaugino and scalar masses will not be achieved without some tuning.

\subsection{Aknowladgements}

I would like to thank Joerg Jaeckel and Valya Khoze for helpfull discussions and advice. This work was supported by an FCT PhD studentship.

\nocite{*}
\bibliographystyle{unsrt}
\bibliography{bibliography}

\begin{thebibliography}{10}

\bibitem{Intriligator:2006dd}
Kenneth~A. Intriligator, Nathan Seiberg, and David Shih.
\newblock {Dynamical SUSY breaking in meta-stable vacua}.
\newblock {\em JHEP}, 04:021, 2006.

\bibitem{Murayama:2007fe}
Hitoshi Murayama and Yasunori Nomura.
\newblock {Simple scheme for gauge mediation}.
\newblock {\em Phys. Rev.}, D75:095011, 2007.

\bibitem{Murayama:2006yf}
Hitoshi Murayama and Yasunori Nomura.
\newblock {Gauge mediation simplified}.
\newblock {\em Phys. Rev. Lett.}, 98:151803, 2007.

\bibitem{Abel:2007jx}
Steven Abel, Callum Durnford, Joerg Jaeckel, and Valentin~V. Khoze.
\newblock {Dynamical breaking of $U(1)_{R}$ and supersymmetry in a metastable
  vacuum}.
\newblock {\em Phys. Lett.}, B661:201--209, 2008.

\bibitem{Abel:2007nr}
Steven~A. Abel, Callum Durnford, Joerg Jaeckel, and Valentin~V. Khoze.
\newblock {Patterns of Gauge Mediation in Metastable SUSY Breaking}.
\newblock {\em JHEP}, 02:074, 2008.

\bibitem{Abel:2008gv}
Steven Abel, Joerg Jaeckel, Valentin~V. Khoze, and Luis Matos.
\newblock {On the Diversity of Gauge Mediation: Footprints of Dynamical SUSY
  Breaking}.
\newblock {\em JHEP}, 03:017, 2009.

\bibitem{Komargodski:2009jf}
Zohar Komargodski and David Shih.
\newblock {Notes on SUSY and R-Symmetry Breaking in Wess-Zumino Models}.
\newblock {\em JHEP}, 04:093, 2009.

\bibitem{Giveon:2009yu}
Amit Giveon, Andrey Katz, and Zohar Komargodski.
\newblock {Uplifted Metastable Vacua and Gauge Mediation in SQCD}.
\newblock 2009.

\bibitem{Abel:2009ze}
Steven~A. Abel, Joerg Jaeckel, and Valentin~V. Khoze.
\newblock {Gaugino versus Sfermion Masses in Gauge Mediation}.
\newblock 2009.

\bibitem{Seiberg:2008qj}
Nathan Seiberg, Tomer Volansky, and Brian Wecht.
\newblock {Semi-direct Gauge Mediation}.
\newblock {\em JHEP}, 11:004, 2008.

\bibitem{Elvang:2009gk}
Henriette Elvang and Brian Wecht.
\newblock {Semi-Direct Gauge Mediation with the 4-1 Model}.
\newblock 2009.

\bibitem{Raby:1998bg}
Stuart Raby and Kazuhiro Tobe.
\newblock {Dynamical SUSY breaking with a hybrid messenger sector}.
\newblock {\em Phys. Lett.}, B437:337--343, 1998.

\bibitem{Dienes:2008rg}
Keith~R. Dienes and Brooks Thomas.
\newblock {New Non-Trivial Vacuum Structures in Supersymmetric Field Theories}.
\newblock {\em AIP Conf. Proc.}, 1116:391--396, 2009.

\bibitem{Ibe:2009bh}
M.~Ibe, Izawa~K. I., and Y.~Nakai.
\newblock {Studying Gaugino Mass in Semi-Direct Gauge Mediation}.
\newblock 2009.

\bibitem{Komargodski:2009pc}
Zohar Komargodski and Nathan Seiberg.
\newblock {Comments on the Fayet-Iliopoulos Term in Field Theory and
  Supergravity}.
\newblock 2009.

\bibitem{1987NuPhB.289..589D}
M.~{Dine}, N.~{Seiberg}, and E.~{Witten}.
\newblock {Fayet-Iliopoulos terms in string theory}.
\newblock {\em Nuclear Physics B}, 289:589--598, 1987.

\bibitem{Abel:2006my}
Steven~A. Abel, Joerg Jaeckel, and Valentin~V. Khoze.
\newblock {Why the early universe preferred the non-supersymmetric vacuum. II}.
\newblock {\em JHEP}, 01:015, 2007.

\bibitem{Katz:2009gh}
Andrey Katz.
\newblock {On the Thermal History of Calculable Gauge Mediation}.
\newblock 2009.

\bibitem{Izawa:1996pk}
Ken-Iti Izawa and Tsutomu Yanagida.
\newblock {Dynamical Supersymmetry Breaking in Vector-like Gauge Theories}.
\newblock {\em Prog. Theor. Phys.}, 95:829--830, 1996.

\bibitem{Intriligator:1996pu}
Kenneth~A. Intriligator and Scott~D. Thomas.
\newblock {Dynamical Supersymmetry Breaking on Quantum Moduli Spaces}.
\newblock {\em Nucl. Phys.}, B473:121--142, 1996.

\bibitem{Ray:2006wk}
Sebastien Ray.
\newblock {Some properties of meta-stable supersymmetry-breaking vacua in
  Wess-Zumino models}.
\newblock {\em Phys. Lett.}, B642:137--141, 2006.

\bibitem{Wess:1992}
Jonathan~Bagger Julius~Wess.
\newblock {\em {Supersymmetry and Supergravity}}.

\bibitem{Dine:2006xt}
Michael Dine and John Mason.
\newblock {Gauge mediation in metastable vacua}.
\newblock {\em Phys. Rev.}, D77:016005, 2008.

\bibitem{Intriligator:2007py}
Kenneth~A. Intriligator, Nathan Seiberg, and David Shih.
\newblock {Supersymmetry Breaking, R-Symmetry Breaking and Metastable Vacua}.
\newblock {\em JHEP}, 07:017, 2007.

\bibitem{RevModPhys.72.25}
Yael Shadmi and Yuri Shirman.
\newblock Dynamical supersymmetry breaking.
\newblock {\em Rev. Mod. Phys.}, 72(1):25--64, Jan 2000.

\bibitem{Giudice:1998bp}
G.~F. Giudice and R.~Rattazzi.
\newblock {Theories with gauge-mediated supersymmetry breaking}.
\newblock {\em Phys. Rept.}, 322:419--499, 1999.

\bibitem{Dine:2009sw}
Michael Dine and John Kehayias.
\newblock {Discrete R Symmetries and Low Energy Supersymmetry}.
\newblock 2009.

\bibitem{Koschade:2009qu}
Daniel Koschade, Moritz McGarrie, and Steven Thomas.
\newblock {Direct Mediation and Metastable Supersymmetry Breaking for SO(10)}.
\newblock 2009.

\end{thebibliography}

\end{document}